\documentclass[aps, prd, 
twocolumn, 
showpacs, showkeys, 
nofootinbib,
superscriptaddress 
]{revtex4-1}
\usepackage[utf8]{inputenc}
\usepackage[T1]{fontenc}
\usepackage{lmodern}
\usepackage{microtype}
\usepackage[english]{babel}

\usepackage{amssymb,amsbsy,amsfonts,amsthm}
\usepackage{amsmath,mathtools}
\allowdisplaybreaks
\usepackage{yhmath} 

\usepackage{slashed}
\usepackage{bm}
\usepackage{times}

\usepackage{soul}

\usepackage[dvipsnames,table,xcdraw]{xcolor}
\definecolor{lightyellow}{RGB}{255,250,205}

\usepackage{graphicx}
\usepackage{epstopdf}
\usepackage{tabularx}
\usepackage{multirow}
\usepackage{longtable}

\usepackage{hyperref}
\hypersetup{
  colorlinks=true,
  linkcolor=blue,  
  citecolor=cyan,
  urlcolor=red,           
}
\usepackage{url}
\usepackage{comment}

\def\orcid#1{\kern .08em\href{https://orcid.org/#1}{\includegraphics[keepaspectratio,width=0.7em]{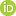}}}

\makeatletter
\newcommand{\vast}{\bBigg@{3}}
\makeatother

\begin{document}

\title{Phenomenological model for the $\gamma\gamma \to \pi^+\pi^-\pi^0$ reaction}

\author{Xiu-Lei Ren \orcid{0000-0002-5138-7415}}
\affiliation{Institut f\"ur Kernphysik \& PRISMA$^+$  Cluster of Excellence, Johannes Gutenberg Universit\"at,  D-55099 Mainz, Germany}
\affiliation{Helmholtz Institut Mainz, D-55099 Mainz, Germany}

\author{Igor Danilkin \orcid{}}
\affiliation{Institut f\"ur Kernphysik \& PRISMA$^+$  Cluster of Excellence, Johannes Gutenberg Universit\"at,  D-55099 Mainz, Germany}

\author{Marc Vanderhaeghen \orcid{0000-0003-2363-5124}}
\affiliation{Institut f\"ur Kernphysik \& PRISMA$^+$  Cluster of Excellence, Johannes Gutenberg Universit\"at,  D-55099 Mainz, Germany}

\begin{abstract}
We present a phenomenological model for the $\gamma\gamma \to \pi^+\pi^-\pi^0$ reaction by including the production mechanism of the $a_2(1320)$ resonance, as well as the contributions from the $\sigma/f_0(500)\, \pi^0$ and $f_2(1270)\, \pi^0 $ production channels. Furthermore, the $\gamma\gamma\to\rho^{\pm} \pi^{\mp}\to\pi^{+}\pi^-\pi^0$ channel, which is essential for a description in the low-energy region, is investigated carefully by introducing the complete set of gauge invariant and Lorentz-covariant tensors for the $\gamma\gamma\to\rho^\pm \pi^\mp$ subprocess. The full amplitude is constructed to yield a correct high-energy Regge behavior. Within our model, we achieve a very reasonable description of ARGUS and L3 data of the total cross section, as well as of the $\pi^\pm\pi^0$ and $\pi^+\pi^-$ invariant mass distributions. We also predict the invariant mass distributions in the $\gamma\gamma$ center-of-mass energy range from $0.8$ GeV to $2.0$ GeV, which will be studied by the forthcoming data of the BESIII collaboration. 
\end{abstract}


\maketitle

\date{\today}

\section{Introduction}
With the precision measurement of the anomalous magnetic moment of the muon, $a_\mu=(g-2)/2$, released by the Muon $g-2$ collaboration~\cite{Muong-2:2021ojo} at Fermi National Accelerator Laboratory (Fermilab) and combined with the measurement of Brookhaven National Laboratory experiment~\cite{Muong-2:2006rrc}, a discrepancy of 4.2~$\sigma$ is found in comparison with the current Standard Model (SM) prediction~\cite{Aoyama:2020ynm}. To meet the accuracy ($\sim$ 1 part-per-million) of the ongoing Fermilab experiment, further efforts are needed to bring down the theoretical error of the SM value. One of the major sources of uncertainty is the hadronic light-by-light (HLbL) scattering contribution. The contributions due to the single pole, the one-loop box diagram, the two-particle cuts, are all well studied, especially for pions,~e.g.~\cite{Bijnens:1995xf,Hayakawa:1997rq,Knecht:2001qf,Colangelo:2017fiz,Hoferichter:2018kwz,Danilkin:2021icn}. Contributions beyond that, in particular the three-pion intermediate states and higher ones are required to achieve a good control of the uncertainty, as stressed in the recent  review by the Muon $g-2$ Theory Initiative~\cite{Aoyama:2020ynm}. 

Towards this goal, a next step is to investigate the two-photon fusion to three pions in detail. This will eventually pave the way towards estimating the  four-point contributions to HLbL with the three pion intermediate states. From the experimental side,  the existing data for the  $\gamma\gamma\to \pi \pi \pi$ process are rather old and have low statistics. Two early experimental investigations of the $\gamma\gamma\to \pi^+\pi^-\pi^0$ reaction were performed by the ARGUS and the L3 Collaborations around 25 years ago~\cite{ARGUS:1996ith,L3:1997mpi}.
An updated analysis of L3 data was carried out in Ref.~\cite{Shchegelsky:2006es}. A comparison between the updated L3 and the ARGUS cross section data shows a significant difference in the low-energy region, which is the most relevant for the HLbL contribution to $a_\mu$. The prospect of new data from the BESIII experiment motivates a renewed interest in this reaction~\cite{Danilkin:2019mhd,Redmer:2019zzr}.

On the theoretical side, the studies of the $\gamma\gamma\to \pi^+ \pi^- \pi^0$ process are also limited. Based on the Current Algebra and the linear sigma model, the $\gamma\gamma\to \pi^+ \pi^- \pi^0$ amplitude has been investigated at lowest order~\cite{Adler:1971nq,Wong:1971ff,Aviv:1971hq,Pratap:1972qd} in the 1970s. With the chiral perturbation theory (ChPT) founded as a powerful tool to describe processes involving low-energy pions, Bos {\it et al.}~\cite{Bos:1994yw} applied ChPT to estimate the total cross section for the $\gamma\gamma\to \pi^+ \pi^- \pi^0$ process at tree-level. Subsequently, the one-loop calculation was carried out by Talavera {\it et al.}~\cite{Talavera:1995fx} and extended within the ``so-called'' generalized ChPT by incorporating the quark condensate in Ref.~\cite{Ametller:1999uz}. 

However, those studies focused  on the very low-energy region, nearby the 3$\pi$ threshold of the two-photon fusion reaction. A phenomenological analysis of $\gamma\gamma\to \pi^+\pi^-\pi^0$ reaction of the experimental data of L3 and ARGUS, covering the low- and intermediate-energy regions, is still missing. Facing the ongoing BESIII experiment, we develop a theoretical model of the $\gamma\gamma\to\pi^+\pi^-\pi^0$ process by accounting for the contributions of the $a_2(1320)$ resonance, as well as the $\sigma/f_0(500)\,\pi^0$, $ f_2(1270)\,\pi^0$, and $\rho^{\pm}(770)\,\pi^\mp$ production channels. 
As a result, we obtain a description for both the total cross section and the invariant mass distributions of ARGUS and L3 experiments \cite{ARGUS:1996ith,L3:1997mpi,Shchegelsky:2006es} with a single parameter to be fixed. The model is able to provide predictions for the forthcoming BESIII data, and can serve as a starting point for further improvements once new data are available.

Our paper is organized as follows: In Sect.~\ref{SecII}, we  present the  $\gamma\gamma\to\pi^+\pi^-\pi^0$ amplitude in our phenomenological model. Within this approach, the existing data of ARGUS and L3 are described  and predictions for the energy range of the forthcoming BESIII data are given in Sect.~\ref{SecIII}. We summarize the main results in Sect.~\ref{SecIV} with perspectives.    

\begin{figure*}[t]
\includegraphics[width=0.65\textwidth]{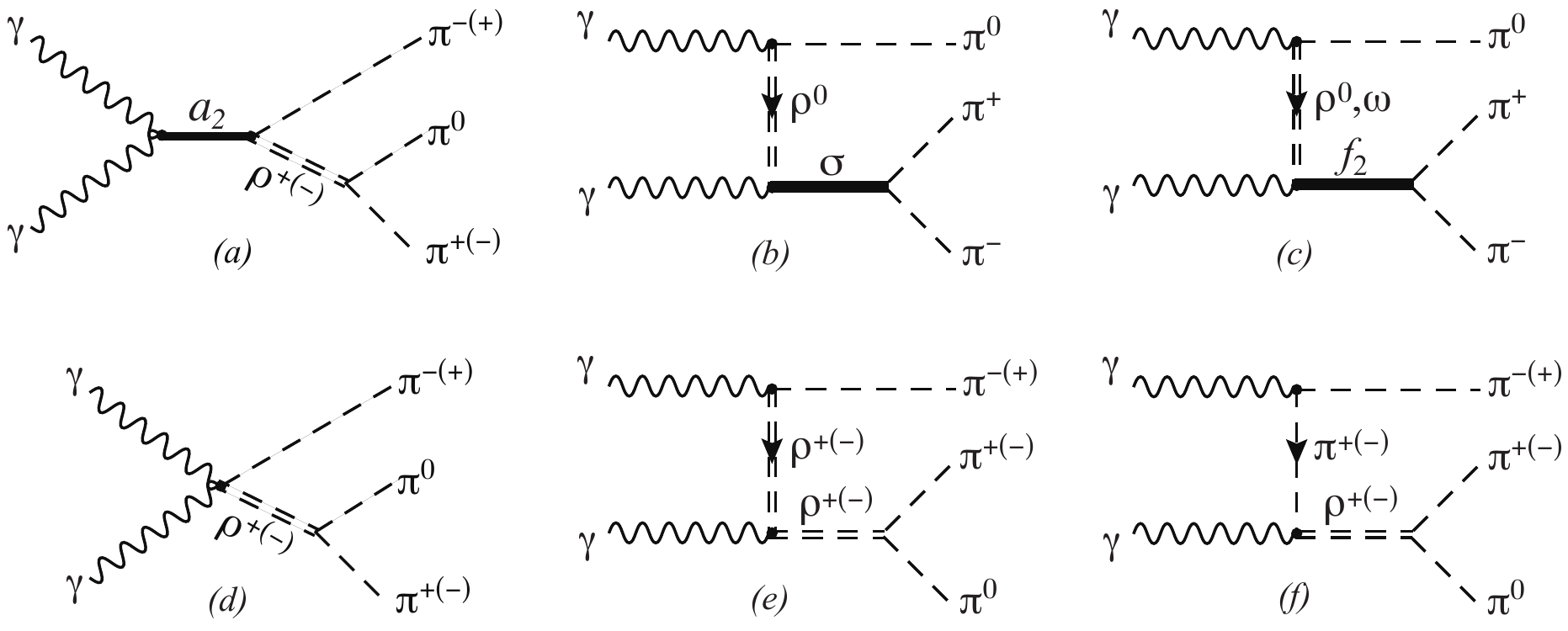}
\caption{Feynman diagrams for the $\gamma\gamma\to\pi^+\pi^-\pi^0$ reaction 
in our model. The diagrams with crossed photon lines are not shown, but included in the calculation.}
\label{Fig:FeynALL}
\end{figure*}

\section{Theoretical framework}~\label{SecII}
In this section, we present the amplitude of the real photon fusion process  $\gamma\gamma\to\pi^+\pi^-\pi^0$, which will be used to describe the current experimental data within the relatively large $\gamma\gamma$ center-of-mass (c.m.) energy range of $0.8$ GeV to $2$ GeV. As shown in Refs.~\cite{ARGUS:1996ith,L3:1997mpi}, this process is dominated by the $a_2(1320)$ resonance.  Besides that, the $\sigma/f_0(500)$, $f_2(1270)$, and $\rho^\pm(770)$  resonances are involved as the intermediate states of quasi two-body production channels due to their strong decay mode to the $\pi \pi$ states.  Therefore, in our model for the $\gamma(k_1)\gamma(k_2) \to \pi^+(p_{\pi^+}) \pi^-(p_{\pi^-}) \pi^0(p_{\pi^0})$ reaction, we parametrize the total amplitude by  several contributing subprocesses: $\gamma\gamma\to a_2(1320) \to\rho^\pm \pi^\mp \to \pi^+\pi^-\pi^0$ resonance production in the $s$-channel, $\gamma\gamma\to \sigma/f_0(500) [f_2(1270)] \pi^0 \to \pi^+\pi^-\pi^0$ production, and $\gamma\gamma\to\rho^\pm \pi^\mp \to \pi^+\pi^-\pi^0$ with the $\rho$ and $\pi$ exchanges in the $t$- and $u$-channel, in order to cover the energy region up to $2$ GeV. The kinematical invariants which we will use in this work in describing the two-photon fusion process are defined as 
\begin{align}
&s = (k_1+k_2)^2,\quad t = (k_1-p_{\pi^+}-p_{\pi^0})^2, \nonumber \\
&u = (k_2-p_{\pi^+}-p_{\pi^0})^2,\quad M^2_{\pi^+\pi^-}= (p_{\pi^+}+p_{\pi^-})^2, \nonumber \\
&M^2_{\pi^+\pi^0}= (p_{\pi^+}+p_{\pi^0})^2,\quad M^2_{\pi^-\pi^0}= (p_{\pi^-}+p_{\pi^0})^2\,. 
\label{eq:mandelstam}
\end{align}
We denote the total amplitude of $\gamma\gamma\to\pi^+\pi^-\pi^0$ as  
\begin{equation}\label{Eq:totalAmp}
\mathcal{M}_{\gamma\gamma\to \pi^+\pi^0\pi^-}  = \mathcal{M}^{a_2} +\mathcal{M}^{f_2} + \mathcal{M}^{\sigma}  + \mathcal{M}^{\rho\pi},
\end{equation}
where each contribution is described in the following subsections and represented by the corresponding Feynman diagrams in Fig.~\ref{Fig:FeynALL}.

\subsection{$\gamma\gamma\to a_2(1320) \to\rho^\pm \pi^\mp \to \pi^+\pi^-\pi^0$ channel}
The $s$-channel contribution of the $a_2(1320)$ resonance production, as the dominant feature of the $\gamma\gamma\to\pi^+\pi^-\pi^0$ reaction, via the $\rho\pi$ decay in Fig.~\ref{Fig:FeynALL}(a) is displayed first. 
Assuming that the $a_2(1320)$ resonance is predominantly produced in a state with helicity-2,
the effective Lagrangian for the $\gamma\gamma\to a_2(1320)$ amplitude can be written as \cite{Drechsel:1999rf}
\begin{equation}
    \mathcal{L}_{\gamma\gamma a_2} = e^2\, \frac{g_{a_2\gamma\gamma}}{m_{a_2}}\, \Phi_{\mu\nu}\, F^{\mu\lambda} F_\lambda^{~\nu},
\end{equation}
where $\Phi_{\mu\nu}$ is the symmetric and traceless tensor describing the spin-2 field, and $F^{\mu\nu}=\partial^\mu A^\nu - \partial^\nu A^\mu$ is the electromagnetic tensor.
The dimensionless coupling $g_{a_2\gamma\gamma}$ is  determined from the two photon decay width
\begin{equation}
\Gamma_{a_2\to\gamma\gamma} = \frac{\pi \alpha^2}{5}\, g_{a_2\gamma\gamma}^2\,m_{a_2},
\end{equation}
where $\alpha=e^2/(4\pi)$ denotes the fine-structure constant. Using the experimental value $\Gamma_{a_2\to\gamma\gamma}^\mathrm{exp}=1.00 \pm 0.06$~ keV~from PDG~\cite{ParticleDataGroup:2022pth}, the absolute value of $g_{a_2\gamma\gamma}$ is fixed as $|g_{a_2\gamma\gamma}|=0.151\pm 0.005$. Note that the PDG average value of $\Gamma_{a_2\to\gamma\gamma}$ are dominated by the results of ARGUS and L3 shown in this paper.   

The effective Lagrangian of $a_2(1320)$ decay to $\rho\pi$ is written as~\cite{Giacosa:2005bw},
\begin{equation}\label{Eq:a2rhopi}
\begin{aligned}
	\mathcal{L}_{a_2\rho\pi} &= \frac{g_{a_2\rho\pi}}{\sqrt{2}}\, \epsilon_{\mu\nu\lambda\sigma}\,(\partial^\mu \bm{\Phi}^{\nu\alpha}-\partial^\nu \bm{\Phi}^{\mu\alpha}) \\
	&\quad \cdot 
	\left[ \partial_\alpha \bm{\pi} \times (\partial^\lambda \bm{\rho}^\sigma - \partial^\sigma \bm{\rho}^\lambda)\right],
\end{aligned}
\end{equation}
where $\bm{\Phi}$, $\bm{\pi}$, $\bm{\rho}$ stand for the isovector $a_2(1320)$, $\pi$, and $\rho$ fields, respectively. From Eq. (\ref{Eq:a2rhopi}) we can calculate the decay width for $a_2\to\rho\pi$
\begin{equation}
  \Gamma_{a_2\to\rho\pi} = \frac{4}{5\pi}\, g_{a_2\rho\pi}^2\left[\frac{\lambda(m_{a_2}^2,m_\rho^2,m_\pi^2)}{4m_{a_2}^2}\right]^{5/2},
\end{equation}
with $\lambda$ being the Gunnar-K\"all\'en function 
\begin{equation}
\lambda(x, y, z)\equiv x^2 + y^2 + z^2 - 2 x y - 2 x z - 2 y z.	
\end{equation}
The coupling $g_{a_2\rho\pi}$ is estimated via the partial decay width of $\Gamma_{a_2\to \rho\pi}=75.0\pm 4.5$ MeV, which is obtained by assuming that the branch ratio $B(a_2\to 3\pi)=70.1\pm 2.7 \%$~\cite{ParticleDataGroup:2022pth} is only from $a_2\to \rho(770)\pi$ decay channel, similar to Ref.~\cite{Belle:2009xpa}. As a result, one obtains the value: $|g_{a_2\rho\pi}|=4.9\pm 0.2$.

The Lagrangian of the rho-meson decay into two pions is given by
\begin{equation}
\mathcal{L}_{\rho\pi\pi}=  g_{\rho\pi\pi}\,(\bm{\pi}\times \partial^\mu \bm{\pi}) \cdot \bm{\rho}_\mu,	
\end{equation}
with the coupling $g_{\rho\pi\pi}=5.97$, which is fixed by the corresponding decay width $\Gamma_{\rho\pi\pi}=149$ MeV, as the isospin average value.

Combining the vertices above, the $s$-channel amplitude of $\gamma(k_1)\gamma(k_2)\to \pi^+(p_{\pi^+})\pi^-(p_\pi^-)\pi^0(p_{\pi^0})$ via the $a_2(1320)$ resonance is given by 
\begin{equation}
  \mathcal{M}^{a_2} = \mathcal{M}_{\rho^+\pi^-}^{a_2} + \mathcal{M}_{\rho^-\pi^+}^{a_2},
\end{equation}
where 
\begin{equation}\label{Eq:a2(1320)}
\begin{aligned}
\mathcal{M}_{\rho^+\pi^-}^{a_2} &= \frac{\sqrt{2}\,e^2}{m_{a_2}}  g_{a_2\gamma\gamma}\,  g_{a_2\rho\pi}\,g_{\rho\pi\pi}  \bigl[k_1^\mu \varepsilon^\lambda(k_1,\lambda_1) - k_1^\lambda \varepsilon^\mu(k_1,\lambda_1) \bigr]\\
&  
 \times \bigl[(k_{2})_\lambda \varepsilon^\nu(k_2,\lambda_2) - (k_2)^\nu \varepsilon_\lambda(k_2,\lambda_2) \bigr] \\
& \times \frac{ \epsilon_{\delta \omega \xi \eta} \, \biggl[ P^\delta \, \Lambda^{\omega\alpha,\mu\nu}(P,\lambda_{a_2}) - (\delta \leftrightarrow \omega) \biggr]}{s-m_{a_2}^2 + i\,m_{a_2}\, \Gamma_{a_2}(s)} \\
& \times \sqrt{\frac{D_2[q_{a_2\to\gamma\gamma}(s)R_{a_2}]}{D_2[q_{a_2\to\gamma\gamma}(m_{a_2}^2)R_{a_2}]}} \sqrt{\frac{D_2[q_{a_2\to\rho\pi}(s)R_{a_2}]}{D_2[q_{a_2\to\rho\pi}(m_{a_2}^2)R_{a_2}]}} \\ 
&  \times 
\frac{\bigl(-g^{\beta \eta} (p_{\rho^+})^\xi + g^{\beta \xi} (p_{\rho^+})^\eta \bigr) \, ({p_{\pi^-}})_{\alpha} \, (p_{\pi^0}-p_{\pi^+})_\beta }{p_{\rho^+}^2 - m_\rho^2 + i\,m_\rho\, \Gamma_\rho(p_{\rho^+}^2)} \\
&\times \frac{D_1[q_{\rho\to\pi\pi}(p_{\rho^+}^2)R_{\rho}]}{D_1[q_{\rho\to\pi\pi}(m_{\rho}^2)R_\rho]}  ,\\[0.5em] 
\mathcal{M}_{\rho^-\pi^+}^{a_2} &= \left. \mathcal{M}_{\rho^+\pi^-}^{a_2}\right|_{p_{\pi^+} \leftrightarrow p_{\pi^-}},
\end{aligned}
\end{equation}	
with the $a_2(1320)$ momentum $P\equiv k_1+k_2$ and the $\rho^+$ momentum $p_{\rho^+}=p_{\pi^+}+p_{\pi^0}$. In Eq.~(\ref{Eq:a2(1320)}),
$\varepsilon^\mu(k_i,\lambda_i)$ denotes the polarization vector of the incoming photons with helicity $\lambda_i$.
The sum over the helicities of $a_2(1320)$ is denoted by \cite{Spehler:1991yw}
\begin{align}\label{Eq:sumspin2}
	\Lambda^{\omega\alpha,\mu\nu}(P,\lambda_{a_2})&\equiv \sum\limits_{\lambda_{a_2}} \varepsilon^{\omega\alpha}(P,\lambda_{a_2})\varepsilon^{*\mu\nu}(P,\lambda_{a_2}) \\
	&= \frac{1}{2} \bigl(K^{\omega\mu}K^{\alpha\nu} + K^{\omega\nu}K^{\alpha\mu}\bigr) - \frac{1}{3}\,K^{\omega\alpha}K^{\mu\nu},\nonumber \\
	K^{\mu\nu}&=-g^{\mu\nu}+\frac{P^\mu P^\nu}{P^2}\,,\nonumber
\end{align}
where $\varepsilon^{\mu\nu}(P,\lambda_{a_2})$ is the spin-2 polarization tensor with helicity $\lambda_{a_2}$. 
The Blatt-Weisskopf form factors $D_\ell(x)$~\cite{Blatt:1952ije} are taken into account in the amplitude for resonances which decay in channels with $\ell \neq0$. For $P$ and $D$-waves, they are given by $D_1(x)=1/(x^2+1)$ and $D_2(x)=1/(x^4+3x^2+9)$. The constants $R_{a_2, \rho}$ are the effective interaction radii (range parameters), which we fixed to  $R_{a_2}=3.1~\mathrm{GeV}^{-1}$~\cite{Belle:2009xpa} and $R_{\rho}=5.3~\mathrm{GeV}^{-1}$~\cite{ParticleDataGroup:2022pth}. Correspondingly, the energy dependent widths of the intermediate $a_2(1320)$ and $\rho^\pm$ resonances appearing in the propagators are 
\begin{equation}
\begin{aligned}
\Gamma_{a_{2}}(s) &=\Gamma_{a_{2}}(m_{a_2} ^2) \Biggl\{ \mathcal{B}(a_2\to \rho\pi) \, \frac{m_{a_2}}{\sqrt{s}}\left[\frac{q_{a_2\to\rho\pi}(s)}{q_{a_2\to \rho\pi}(m_{a_2}^2)}\right]^{5} \\ 
&\quad \times  \frac{D_2[q_{a_2\to\rho\pi}(s)R_{a_2}]}{D_2[q_{a_2\to\rho\pi}(m_{a_2}^2)R_{a_2}]} \, \Theta\left(s-\left(m_{\pi}+ p_{\rho^+}\right)^2\right) \\
& \quad + \eta\pi, K\bar{K}, \omega \pi\pi~\mathrm{channels} \Biggr\} ,\\ 
\Gamma_{\rho}(p_{\rho^+}^2)&=\Gamma_{\rho}(m_{\rho}^2) \, \frac{m_\rho}{\sqrt{p_{\rho^+}^2}} \left[\frac{q_{\rho\to\pi\pi}(p_{\rho^+}^2)}{q_{\rho\to\pi\pi}(m_\rho^2)}\right]^{3} \\
&\quad  \times \frac{D_1[q_{\rho\to\pi\pi}(p_{\rho^+}^2)R_\rho]}{D_1[q_{\rho\to\pi\pi}(m_{\rho}^2)R_\rho]} \Theta\left(p_{\rho^+}^2-4 m_{\pi}^{2}\right)\,.
\end{aligned}
\end{equation}
The explicit forms of the subdominant $\eta\pi$, $K\bar{K}$, $\omega \pi\pi$ contributions to the $a_2(1320)$ width can be found in Ref.~\cite{Belle:2009xpa}.
The rest frame momenta of the considered channels are written as 
\begin{equation}
\begin{aligned}
    q_{a_2\to \gamma\gamma}(s) &= \frac{\sqrt{s}}{2},\quad 
	q_{a_2\to \rho\pi}(s) = \frac{\lambda^{1/2}(s, p_{\rho^+}^2, m_\pi^2)}{2\sqrt{s}}, \\
	q_{\rho\to\pi\pi}(p_{\rho^+}^2) &= \frac{\lambda^{1/2}(p_{\rho^+}^2, m_\pi^2, m_\pi^2)}{2\sqrt{p^2_{\rho^+}}}.
\end{aligned}
\end{equation}

\subsection{$\gamma\gamma\to \sigma/f_0(500) \, \pi^0 \to \pi^+\pi^-\pi^0$ channel}

\begin{figure}[t]
\includegraphics[width=0.20\textwidth]{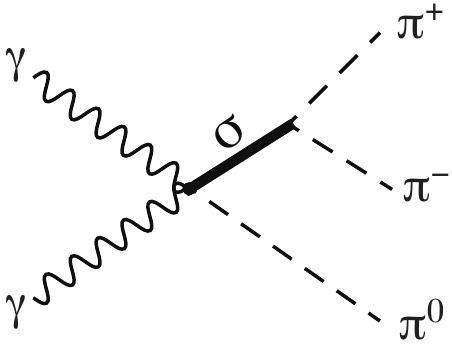}
  \caption{Contact term corresponding to the $\sigma$ contribution. The bold solid line represents the quasi two-pion state, which is described by the Omn\`{e}s function, as explained in the text.}
  \label{Fig:sigma_cont}
\end{figure}
Besides the dominant $a_2(1320)$ production channel, we need to account for the contributions to the $\gamma\gamma\to\pi^+\pi^-\pi^0$ process in the low-energy region. The effective Lagrangian which describes the ``sigma''~\footnote{Note that the ``sigma" here stands for the contact interaction with production of $\pi^+\pi^-$ in a $J=I=0$ state.} contribution to $\gamma\gamma\to\pi^+\pi^-\pi^0$ in the linear sigma model is given by~\cite{Adler:1971nq}
\begin{equation}\label{Eq:Lsigma}
  \mathcal{L}_\text{$\sigma$-model} =\frac{e^2}{16}\,F^\mathrm{WZW}_{3\pi}\, \epsilon_{\mu \nu \alpha \beta} \, F^{\mu \nu} \, F^{\alpha \beta} \, \pi^{0}\, \left(\bm{\pi} \cdot \bm{\pi}\right),
\end{equation} 
where $F^\mathrm{WZW}_{3\pi}\equiv 1/(4\,\pi^2 f_\pi^3)$ with the pion decay constant $f_\pi=92.4$ MeV. 
The chiral contact amplitude is then given  by 
\begin{equation}\label{Eq:sigmacont}
\begin{aligned}
    \mathcal{M}_\text{$\sigma$-model} = & i\,e^2 F_{3\pi}^\mathrm{WZW}\, \epsilon_{\mu\nu\alpha\beta}\,  k_1^{\mu}\,k_2^{\nu}\, \varepsilon^\alpha(k_1,\lambda_1)\,\varepsilon^\beta(k_2,\lambda_2). 
\end{aligned}	
\end{equation}
The contact term in the effective field theory can be thought of as the heavy vector meson exchange as shown in Fig.~\ref{Fig:FeynALL}(b). Therefore, including one effective vector meson mass, which we approximate by $m_\rho$, the amplitude, corresponding with  Fig.~\ref{Fig:sigma_cont}, can be generalized to 
\begin{equation}
\begin{aligned}
    \mathcal{M}^{\sigma} &= i\,e^2\,F_{3\pi}(t_\sigma, u_\sigma)\, \epsilon_{\mu\nu\alpha\beta}\,  \\
    &\quad \times k_1^{\mu}\,k_2^{\nu}\, \varepsilon^\alpha(k_1,\lambda_1)\,\varepsilon^\beta(k_2,\lambda_2)\, \Omega(M_{\pi^+\pi^-}^2),
\end{aligned}
\end{equation}
where the $F_{3\pi}$ is parameterized as 
\begin{equation}\label{Eq:f3pion}
    F_{3\pi}(t_\sigma, u_\sigma) = F_{3\pi}^\mathrm{WZW}\left[-\frac{m_\rho^2}{2} \biggl( \frac{1}{t_\sigma - m_\rho^2} +\frac{1}{u_\sigma - m_\rho^2} \biggr) \right],
\end{equation}
with $t_\sigma=(k_1-p_{\pi^0})^2$ and $u_\sigma=(k_2-p_{\pi^0})^2$. Note that $\mathcal{M}^{\sigma}$ amplitude reduces to the contact term in the limit $m_\rho\to \infty$. Besides, in the above amplitude, we considered the rescattering effect of the final $\pi^+\pi^-$ by including the S-wave isospin $I=0$ Omn\`{e}s function,
\begin{equation}
\Omega(x)=\exp\left\{\frac{x}{\pi}\int_{4m_\pi^2}^{\infty}\frac{d x'}{x'}\frac{\delta(x')}{x'-x}\right\}\,,
\end{equation}
with the elastic phase shift input from \cite{Danilkin:2020pak,Danilkin:2022cnj}. It accounts for the rescattering through the $\sigma/f_0(500)$ resonance. 

Furthermore, to obtain the correct behavior of the amplitude in the high-energy region, one needs to modify the Lagrangian based amplitude. In our approximation, we will use the vertices from the corresponding Lagrangians and obtain the correct high-energy behavior by taking into account the exchange of a Regge trajectory (representing the exchange of a family of particles with the same internal quantum numbers). More specifically, the usual pole-like Feynman propagator of a single particle is replaced by its Reggezied counterpart. The details can be found in Ref.~\cite{Guidal:1997hy}. For instance, for the $\rho$ meson propagator it amounts to the following replacements in Eq.~\eqref{Eq:f3pion},
\begin{equation}\label{Eq:Regge_rho}
    \frac{1}{p^2-m_\rho^2} \rightarrow \mathcal{P}^{\rho}(s, p^2),
\end{equation}
where $p$ stands either for $k_1-p_\pi^0$ ($t$-channel) or for $k_2-p_\pi^0$ ($u$-channel). The Regge propagator of the $\rho$ meson is given by~\cite{Guidal:1997hy}
\begin{equation}
\mathcal{P}^{\rho}(s,x) \equiv \left(\frac{s}{s_{0}}\right)^{\alpha_{\rho}(x)-1} \frac{\pi \alpha_{\rho}^{\prime}}{\sin \left(\pi \alpha_{\rho}(x)\right)} \left( \frac{-1+e^{-i \pi \alpha_{\rho}(x)}}{2 \, \Gamma\left(\alpha_{\rho}(x)\right)}\right),	
\end{equation}
where the Gamma function $\Gamma(\alpha(x))$ ensures that the propagator only has poles in the time-like region. 
The $\rho$ Regge trajectory is given by $\alpha_\rho(x)=0.55 + 0.8 \,x$ and
the mass scale is conventionally taken as $s_0=1$ GeV$^2$. Note that the Regge propagator reduces to the Feynman propagator when approaching the first pole on the  trajectory, i.e. for $p^2\to m_\rho^2$.  

\subsection{$\gamma\gamma\to f_2(1270) \, \pi^0 \to \pi^+\pi^-\pi^0$ channel}
The contribution of $f_2(1270)$ resonance is dominated by the vector-meson left-hand cut in the $t$- and $u$-channel, as shown in Fig.~\ref{Fig:FeynALL}(c). In the following, we explicitly denote the $\rho^0$ exchange contribution, and include the $\omega$ counterpart in the effective coupling. The needed effective Lagrangians for the amplitude of $\gamma\gamma\to f_2(1270) \, \pi^0 \to \pi^+\pi^-\pi^0$ channel are,
\begin{equation}
\begin{aligned}
    \mathcal{L}_{f_2\rho\gamma} &= e\,\frac{g_{f_2\rho\gamma}}{m_{f_2}}\, \Phi_{\mu\nu} \, F^{\mu\lambda} (\partial_{\lambda}\rho^{0,\nu} - \partial^\nu \rho^0_\lambda),\\
	\mathcal{L}_{f_2\pi\pi} & = \frac{g_{f_2\pi\pi}}{m_{f_2}}\, \Phi_{\mu\nu} \, \left(\partial^\mu\bm{\pi} \cdot \partial^\nu\bm{\pi}\right),\\
	\mathcal{L}_{\rho\pi\gamma} & = e\, \frac{g_{\rho\pi\gamma}}{m_\pi}\,\epsilon_{\mu\nu\alpha\beta}\, \partial^\beta A^\mu\, \left(\bm{\pi}\cdot \partial^\alpha \bm{\rho}^\nu\right),
\end{aligned}	
\end{equation}
where the tensor $\Phi_{\mu\nu}$ denotes the $f_2(1270)$ field. 
The dimensionless coupling $g_{f_2\pi\pi}\simeq 23.67$ is fixed by the empirical decay widths of $\Gamma_{f_2\to \pi\pi}= 157.2$ MeV~\cite{ParticleDataGroup:2022pth} by using 
\begin{equation}
    \Gamma_{f_2\to \pi\pi} = \frac{g_{f_2\pi\pi}^2}{1280\,\pi} m_{f_2}\left(1-\frac{4\,m_\pi^2}{m_{f_2}^2}\right)^{5/2}. 
\end{equation}
The coupling $g_{\rho\pi\gamma}$ is determined via the $\rho$ meson decay width $\Gamma_{\rho^0 \to \pi^0 \gamma}$,
\begin{equation}
    \Gamma_{\rho\to\pi \gamma} = \frac{\alpha\, g_{\rho\pi\gamma}^2}{24} \frac{m_\rho^3}{m_\pi^2} \left(1-\frac{m_\pi^2}{m_\rho^2}\right)^3.
\end{equation}
Based on the isospin symmetry, we take $\Gamma_{\rho^0 \to \pi^0 \gamma} \simeq \Gamma_{\rho^\pm \to \pi^\pm \gamma}$~\cite{ParticleDataGroup:2022pth} and fix its value from PDG $\Gamma_{\rho^\pm \to \pi^\pm \gamma}  =68\pm 7$ keV.
The obtained coupling $g_{\rho\pi\gamma}=0.102\pm 0.005$ is also consistent with the value obtained from the extrapolation of the lattice QCD result on $\gamma^{(*)}\pi\to \pi\pi$ Ref.~\cite{Niehus:2021iin,Briceno:2016kkp}.  The coupling $g_{f_2\rho\gamma}$ is an unknown parameter, which will be determined a fit to the total cross section of $\gamma\gamma\to \pi^+\pi^-\pi^0$ in the following. 

Thus, one can obtain the contribution of $f_2(1270)$ resonance to the amplitude of $\gamma(k_1)\gamma(k_2)\to \pi^+(p_{\pi^+})\pi^-(p_{\pi^-})\pi^0(p_{\pi^0})$ in the $t$- and $u$-channel as: 
\begin{equation}
   \mathcal{M}^{f_2} = \mathcal{M}^{f_2}_\text{$t$-ch} + \mathcal{M}^{f_2}_\text{$u$-ch},
\end{equation}
with 
\begin{equation}\label{Eq:Amp_f2}
\begin{aligned}
\mathcal{M}_\text{$t$-ch}^{f_2} & = \frac{2 e^2  }{m_{\pi}\, m_{f_2}^2}\,g_{f_2\pi\pi}g_{f_2\rho\gamma}g_{\rho\pi\gamma} \, \varepsilon^{\mu}(k_1,\lambda_1)\varepsilon^{\nu}(k_2,\lambda_2)\, \\
	& \times \frac{ (p_{\pi^+})_\alpha \, (p_{\pi^-})_\beta\,
\Lambda^{\alpha\beta,\delta\omega}(p_{f_2},\lambda_{f_2})}{p_{f_2}^2-m_{f_2}^2+im_{f_2}\Gamma_{f_2}(p_{f_2}^2)}  \\
& \times  \sqrt{\frac{D_2[q_{f_2\to\rho\gamma}(p_{f_2}^2)R_{f_2}]}{D_2[q_{f_2\to\rho\gamma}(m_{f_2}^2)R_{f_2}]} } \sqrt{\frac{D_2[q_{f_2\to\pi\pi}(p_{f_2}^2)R_{f_2}]}{D_2[q_{f_2\to\pi\pi}(m_{f_2}^2)R_{f_2}]} } \\
& \times  \Biggl[\epsilon_{\mu\kappa \xi \eta} \biggl( ({k_2})_{\delta}\left( g^{~\kappa}_\nu ({p_{\rho^0}})_\omega -  g^{~\kappa}_\omega  ({p_{\rho^0}})_\nu\right) \\
&\qquad  + g_{\nu\delta}\left( g^{~\kappa}_\omega k_2\cdot p_{\rho^0} - (k_2)^\kappa ({p_{\rho^0}})_\omega\right)\biggr)  \frac{(p_{\rho^0})^\xi\, (k_{1})^\eta }{p_{\rho^0}^2-m_\rho^2 } \Biggr],\\
\mathcal{M}_\text{$u$-ch}^{f_2} &=\left. \mathcal{M}_\text{$t$-ch}^{f_2}\right|_{\scriptsize 
\begin{pmatrix}
  k_1 \\
  \mu 
\end{pmatrix} \leftrightarrow 
\begin{pmatrix}
  k_2 \\
  \nu 
\end{pmatrix}
},
\end{aligned}	
\end{equation}
with the $f_2(1270)$ momentum $p_{f_2}=p_{\pi^+}+p_{\pi^-}$, the $\rho^0$ momentum $p_{\rho^0}=k_1-p_{\pi^0}$ for the $t$-channel and $p_{\rho^0}=k_2-p_{\pi^0}$ for the $u$-channel.  The projector $\Lambda^{\alpha\beta,\delta\omega}(p_{f_2},\lambda_{f_2})$ represents the sum over the helicities of $f_2(1270)$, which follows the same definition as given in Eq.~(\ref{Eq:sumspin2}), with 
the replacements $a_2 \leftrightarrow f_2$ and 
$P \leftrightarrow p_{f_2}$.  Note that the energy dependent width $\Gamma_{f_2}(p_{f_2}^2)$ included in the propagator of the $f_2(1270)$ resonance has the following form:
\begin{equation}
\begin{aligned}
\Gamma_{f_{2}}(p_{f_2}^2)&=\Gamma_{f_{2}}(m_{f_2}^2)\, \frac{m_{f_2}}{\sqrt{p_{f_2}^2}}
 \Biggl[\frac{q_{f_2\to\pi\pi}(p_{f_2}^2)}{q_{f_2\to\pi\pi}(m_{f_2}^2)}\Biggr]^{5} \\
 &\quad\times \frac{D_2[q_{f_2\to\pi\pi}(p_{f_2}^2)R_{f_2}]}{D_2[q_{f_2\to\pi\pi}(m_{f_2}^2)R_{f_2}]} \Theta\left( p_{f_2}^2-4\,m_{\pi}^2\right),
\end{aligned}
\end{equation}
where the effective range radius $R_{f_2}=3.6~\mathrm{GeV}^{-1}$ was taken from~\cite{Belle:2007ebm}. The rest frame momenta appearing in the $D_2$ functions are defined as  
\begin{equation}
\begin{aligned}
    q_{f_2\to\rho\gamma}(p_{f_2}^2) &= \frac{\lambda^{1/2}(p_{f_2}^2, (k_1-p_{\pi^0})^2,0)}{2\sqrt{s}}, \\
	q_{f_2\to\pi\pi}(p_{f_2}^2) & = \frac{\lambda^{1/2}(p_{f_2}^2, m_\pi^2, m_\pi^2)}{2\sqrt{p^2_{f_2}}}. 
\end{aligned}
\end{equation}
Furthermore, to obtain the correct behavior in the high-energy region, the $\rho$ meson propagators, in the $t$- and $u$-channel of Eq.~\eqref{Eq:Amp_f2}, are replaced by their Reggezied counterparts using the replacement of  Eq.~\eqref{Eq:Regge_rho}. 

\subsection{$\gamma\gamma\to\rho^\pm \pi^\mp \to \pi^+\pi^-\pi^0$ channel}
For the $\gamma \gamma\to \rho^\pm \pi^\mp \to \pi^+\pi^-\pi^0 $ processes, the Feynman diagrams are given in Fig.~\ref{Fig:FeynALL}(d-g), where the diagrams with crossed photon lines are not shown but are included in the calculation. The relevant effective Lagrangians are  
\begin{equation}\label{Eq:LagRho}
\begin{aligned}
  \mathcal{L}_{\gamma\pi\pi} &= -e\, A_\mu\, \left(\bm{\pi}\times  \,\partial^\mu\bm{\pi}\right)_3, \\
  \mathcal{L}_{\gamma\rho\rho} &= e\,A_\mu \, \bigl( \bm{\rho}_\nu\times \left(\partial^\mu\bm{\rho}^\nu -  \partial^\nu \bm{\rho}^\mu \right)\bigr)_3,\\
  \mathcal{L}_{\gamma\gamma\rho\pi} &= e^2\,\frac{ g_{\rho\pi\gamma}}{m_\pi}\, \epsilon_{\mu\nu\alpha\beta}\,(\partial^\beta A^\mu) A^\alpha \, \left(\bm{\pi} \times \bm{\rho}^\nu \right)_3,
\end{aligned}
\end{equation}
where the symbol $(\bm{X})_3$ denotes the third component of isovector $\bm{X}$. 

The amplitude of the $\gamma \gamma\to \rho^{\pm} \pi^{\mp} \to \pi^+\pi^-\pi^0 $ process can be written as 
\begin{equation}
 \mathcal{M}^{\rho\pi} = \varepsilon^{\mu}(k_1,\lambda_1)\,\varepsilon^{\nu}(k_2,\lambda_2)\,\mathcal{M}_{\mu\nu}^{\rho\pi}.
\end{equation}
The tensor $\mathcal{M}_{\mu\nu}^{\rho\pi}$ is obtained by explicitly separating the whole process into the subprocess $\gamma\gamma\to\rho^\pm\pi^\mp$  with two-body final states and the subprocess of 
$\rho$-meson decay to $\pi\pi$ final states, as:  
\begin{equation}
\begin{aligned}
   \mathcal{M}_{\mu\nu}^{\rho\pi} &= \mathcal{M}_{\mu\nu\alpha}^{\gamma\gamma\to\rho^+\pi^-}
   \frac{g_{\rho\pi\pi} (p_{\pi^0} - p_{\pi^+})_\beta K_{\rho}^{\alpha\beta}(p_{\pi^+}+p_{\pi^0})}{M_{\pi^+\pi^0}^2 - m_\rho^2 + i\, m_\rho\,\Gamma_\rho(M_{\pi^+\pi^0}^2)} \\
   &\quad \times \frac{D_1[q_{\rho\to\pi\pi}(M_{\pi^+\pi^0}^2)R_\rho]}{D_1[q_{\rho\to\pi\pi}(m_{\rho}^2)R_\rho]} \\
    & +  \mathcal{M}_{\mu\nu\alpha}^{\gamma\gamma\to\rho^-\pi^+} \frac{g_{\rho\pi\pi} (p_{\pi^-}-p_{\pi^0})_\beta K_{\rho}^{\alpha\beta}(p_{\pi^-} + p_{\pi^0}) }{M_{\pi^-\pi^0}^2 - m_\rho^2 + i\,m_\rho\,\Gamma_\rho(M_{\pi^-\pi^0}^2)} \\
    &\quad \times \frac{D_1[q_{\rho\to\pi\pi}(M_{\pi^-\pi^0}^2)R_\rho]}{D_1[q_{\rho\to\pi\pi}(m_{\rho}^2)R_\rho]},
\end{aligned}
\end{equation}
with $K_{\rho}^{\alpha\beta}(p)\equiv -g^{\alpha\beta}+ 
   p^\alpha p^\beta / p^2$.
The amplitudes of the subprocess $\gamma\gamma \to \rho^\pm\pi^\mp $ are expressed through the tensor decomposition,
\begin{equation}
 \begin{aligned}
  \mathcal{M}_{\mu\nu,\alpha}^{\gamma\gamma \to \rho^\pm\pi^\mp} &= \sum\limits_{i=1}^6\, T^i_{\mu\nu,\alpha}\left(k_1,k_2;p_{\rho^\pm}-p_{\pi^\mp}\right) \\
   & \times F_i^{ \rho^\pm\pi^\mp}\big(M_{\pi^\pm\pi^0}^2,(k_2-p_{\pi^\mp})^2,(k_1-p_{\pi^\mp})^2\big),
 \end{aligned}
 \end{equation}
where the invariant amplitudes $F_i^{\rho^\pm\pi^\mp}$ are free of kinematic constraints. The complete set of gauge invariant and Lorentz-covariant tensors $T_{\mu\nu,\alpha}^i$ for the $\gamma\gamma \to V P$ process is given in Eq.~\eqref{Eq:tensorbasis} of the Appendix. 
The six scalar functions corresponding to Eq.~(\ref{Eq:LagRho}) are also given in Eqs.~\eqref{Eq:scalarfuns2}-\eqref{Eq:scalarfuns7}. 
 
To extend the above amplitude into the high-energy region, 
we will again assume the amplitude to be dominated by Regge poles, and calculate the residues of the pion and rho Regge exchanges based on the amplitudes calculated with Feynman propagators. This amounts to drop the polynomial term in the scalar amplitude $F_2^{ \rho^\pm\pi^\mp}$, as it does not contribute to the residues, and to replace the  $\rho$ and $\pi$ propagators 
in the scalar amplitudes $F_{1-6}^{ \rho^\pm\pi^\mp}$ by their reggezied counterparts. The Reggeized $\rho$-meson propagator is already given in Eq.~\eqref{Eq:Regge_rho}. In turn, the Reggeized $\pi$ propagator has the following form~\cite{Guidal:1997hy}
\begin{equation}
\frac{1}{p^2 -m_\pi^2} \rightarrow \mathcal{P}^{\pi}(s,p^2), 
\end{equation}
where $p^2$ stands again for the squared momentum transfer of the corresponding $t$- or $u$-channel processes and  
\begin{equation}
\mathcal{P}^{\pi}(s,x)\equiv \left(\frac{s}{s_{0}}\right)^{\alpha_{\pi}(x)} \frac{\pi \alpha_{\pi}^{\prime}}{\sin \left(\pi \alpha_{\pi}(x)\right)} \left(\frac{1+e^{-i \pi \alpha_{\pi}(x)}}{2\, \Gamma\left(1+\alpha_{\pi}(x)\right)}\right),
\end{equation}
where the pion Regge trajectory is given by $\alpha_\pi(x) = 0.7(x-m_\pi^2)$.

It is worthy to point out that the contribution of the Deck mechanism~\cite{Deck:1964hm} via the double-exchange of $\rho $ and $\pi$ mesons in $t$- and $u$-channels is also relevant to the $\gamma\gamma\to\pi^+\pi^-\pi^0$ process. We have evaluated their contributions and found the corresponding effects to be around 10\% or less of the $\gamma\gamma\to\rho\pi$ contribution in the energy region $0.8<W<1.0$ GeV. Facing the current significant uncertainty when comparing the L3 and ARGUS data in the low-energy region (see Fig.~3 in the range $0.8<W<1.0$ GeV), we do not include the contribution of Deck mechanism in the present work. Their effects will be carefully investigated with the more accurate forthcoming data from the BESIII experiment.

\begin{table}[b]
\centering
\caption{Values of resonance ($R$) parameters used in our model.}
\begin{tabular}{ccccccc} 
 \hline\hline
  &  & & $m_R$ [MeV] & $\Gamma_R$ [MeV] &  &  \\ 
  \hline
  &  & &  & &  $g_{a_2\gamma\gamma}$ & $g_{a_2\rho\pi}$ \\ [0.5ex] 
   & $a_2(1320)$ & & $1316.9$ & $105$ &  $0.151$ &  $4.9$  \\
 \hline 
 & & &  & &  $g_{f_2\rho \gamma}$ & $g_{f_2\pi \pi}$  \\ [0.5ex] 
   & $f_2(1270)$ & & $1275.5$ & $186.7$ & $-27.5$  &   $23.67$ \\
   \hline 
  & & &  & &  $g_{\rho \pi \gamma}$ &  $g_{\rho \pi \pi}$ \\ [0.5ex] 
& $\rho(770)$ & & $775$ & $149$ &   $0.102$ &  $5.97$ \\ 
 \hline \hline
\end{tabular}
\label{Tab:couplingvalues}
\end{table}

\begin{figure}[t]
    \centering
    \includegraphics[width=0.47\textwidth]{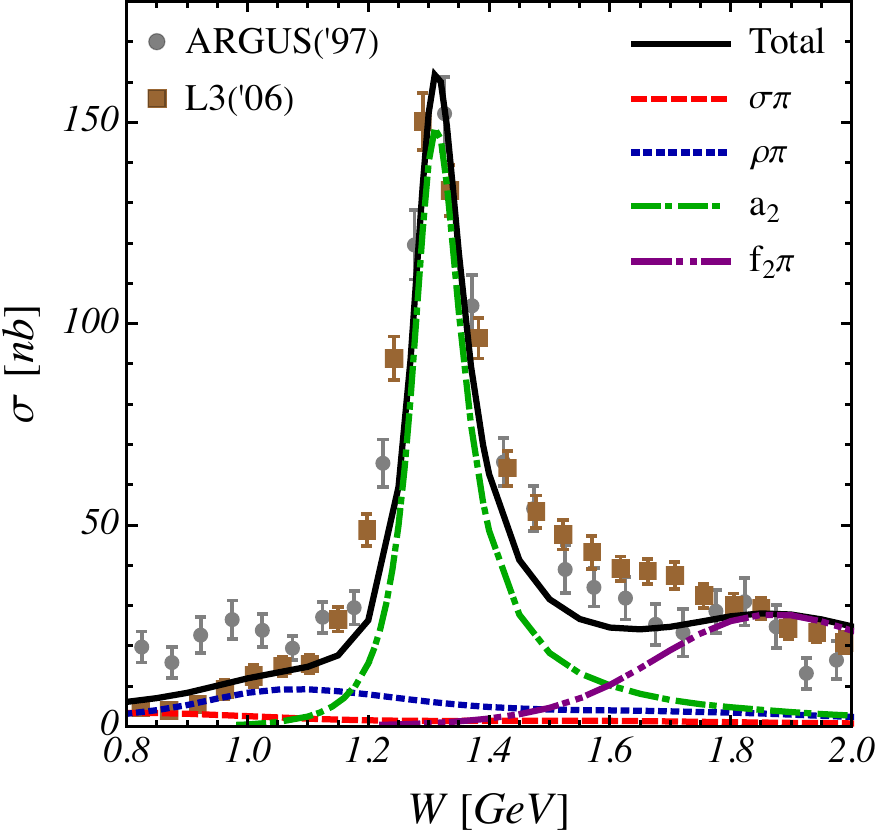}
    \caption{Total cross section for $\gamma\gamma\to\pi^+\pi^-\pi^0$. The gray dots and brown squares are the data points from ARGUS~\cite{ARGUS:1996ith} and L3~\cite{Shchegelsky:2006es}, respectively. The solid line denotes the full results of our model. The different contributions of $\sigma/f_0(500)\pi^0$, $\rho^\pm\pi^\mp$, $a_2(1320)$, and $f_2(1270)\pi^0$ channels are also presented.}
    \label{Fig:TotCross}
\end{figure}

\section{Results and discussion}\label{SecIII}
We are now in the position to describe the experimental observables of the $\gamma\gamma\to\pi^+\pi^-\pi^0$ reaction using the constructed amplitude. The differential cross section for $\gamma(k_1)\gamma(k_2)\to \pi^+(p_{\pi^+})\pi^-(p_\pi^-)\pi^0(p_{\pi^0})$ process is given by
\begin{equation}
\begin{aligned}
   &\frac{d^4\sigma}{ d\,M_{\pi^+\pi^0}^2\,d\,M_{\pi^+\pi^-}^2\, d\,t\, d\phi_{\pi^+}^*} \\
   &= \frac{1}{(2\pi)^4} \frac{\overline{\sum_i} \sum_f |\mathcal{M}_{\gamma\gamma\to \pi^+\pi^-\pi^0}|^2 }{32 s^2 \lambda^{1/2}(s, M_{\pi^+\pi^0}^2, m_\pi^2)} ,  
\end{aligned}
\end{equation} 
where the kinematical invariants are defined in Eq.~(\ref{eq:mandelstam}). In the following, we denote 
the total energy in the $\gamma \gamma$ c.m. frame as $W \equiv \sqrt{s}$. 
The above form is convenient to generate the Dalitz plot and to calculate the  projected invariant mass distributions $d\sigma/dM_{\pi^+\pi^0}$ and $d\sigma/dM_{\pi^+\pi^-}$. 
The solid angle $\Omega_{\pi^+}^* = (\theta_{\pi^+}^*, \phi_{\pi^+}^*)$ is defined  in the rest frame of $\pi^+\pi^0$, with respect to the direction of the $\pi^+\pi^0$ momentum in the $\gamma \gamma$ c.m. frame. 

The average over both photon helicities of the squared amplitude can be written as 
\begin{equation}
\begin{aligned}
  &\overline{\sum_i} \sum_f |\mathcal{M}_{\gamma\gamma\to \pi^+\pi^-\pi^0}|^2 \\
  &\equiv\frac{1}{4} \left( |\mathcal{M}_{++}|^2 + |\mathcal{M}_{+-}|^2  + |\mathcal{M}_{-+}|^2 + |\mathcal{M}_{--}|^2\right),
\end{aligned}
\end{equation} 
where the $\mathcal{M}_{\lambda_1,\lambda_2}$ denotes the helicity amplitude of the photon fusion process. Following the definition of the total amplitude in Eq.~\eqref{Eq:totalAmp}, the $s$-channel $a_2(1320)$ amplitude only contributes to the $\mathcal{M}_{+-}$ and $\mathcal{M}_{-+}$ amplitudes, while the $\sigma/f_0(500)\, \pi^0$ channel has only $\mathcal{M}_{++}$ and $\mathcal{M}_{--}$ components. Therefore, these contributions do not interfere. The other two channels with $f_2(1270)\, \pi^0$ and $\rho^{\pm}\pi^{\mp}$ contribute to all helicity amplitudes. Thus, one needs to specify the relative phases of amplitudes from those four channels. The phases of $\mathcal{M}^{\sigma}$ and $\mathcal{M}^{\rho\pi}$ are determined by reproducing the chiral amplitudes at low energy. While, the relative phases of amplitudes $\mathcal{M}^{a_2}$ and $\mathcal{M}^{f_2}$ are a priori not known. We find a slightly better description of total cross section around $W=1.3$ GeV for the case of the constructive interference between the $\mathcal{M}^{\rho\pi}$ and $\mathcal{M}^{a_2}$, thus fixing the phase of $\mathcal{M}^{a_2}$, which fixes the sign of the product $g_{a_2\gamma\gamma} g_{a_2\rho\pi}$.

To describe the total cross section and the invariant mass distributions, we need to determine the effective couplings in our model. As discussed in Section~\ref{SecII}, most couplings are obtained by reproducing the corresponding decay widths. The only unknown coupling in our description is $g_{f_2\rho\gamma}$, which is determined by reproducing the total cross section $\sigma(W=1.85)\approx 28$ nb, because the ARGUS and L3 data are consistent in this energy region and the contribution of the $f_2(1270)\pi^0$ channel is dominant. Furthermore, the description of the $\pi^+\pi^-$ invariant mass distribution prefers a destructive interference between $\mathcal{M}^{f_2}$ and $\mathcal{M}^{a_2}$, thus fixing the sign of $g_{f_2\rho\gamma}$. 
In Table~\ref{Tab:couplingvalues}, we list the values of couplings and the PDG values of masses and widths of resonances~\cite{ParticleDataGroup:2022pth} as used in our calculation.

\begin{figure}[b]
  \includegraphics[width=0.47\textwidth]{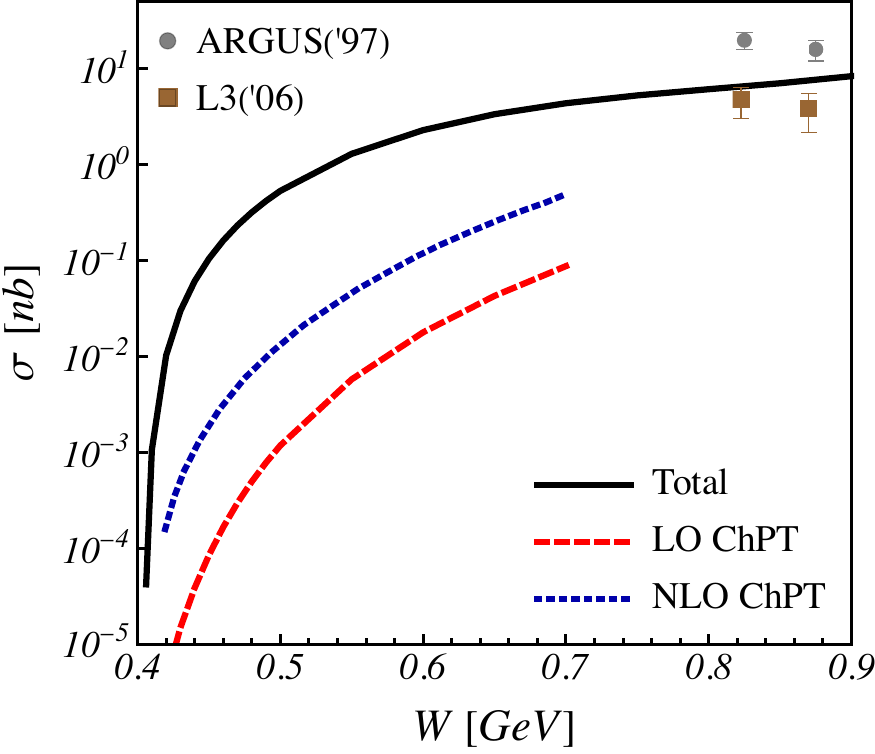}
  \caption{Our prediction of total cross section (black solid line) in the low energy region: $3m_\pi<W<0.9$ GeV. The chiral results of Ref.~\cite{Talavera:1995fx} at LO (red dashed line) and NLO (blue dotted line) are shown.}
\label{Fig:TotCross_lowE}
\end{figure}

First we present the description of the total cross section as compared to the ARGUS and L3 data in Fig.~\ref{Fig:TotCross}. As mentioned in the introduction, the experimental data of ARGUS and L3 show significant differences in the  low-energy region. Our theoretical prediction is consistent with L3 data up to $1.1$ GeV. The unsatisfactory state of the data in the low-energy region will hopefully be resolved by the forthcoming BESIII data. The $\sigma/f_0(500) \pi^0$ contribution dominates around $0.8$ GeV and gradually decreases with increasing energy. Subsequently, the $\rho\pi$ channel starts  contributing at $0.8$ GeV and dominates up to $1.15$ GeV before the effect of the $a_2(1320)$ production in the $s$-channel takes over. In the energy region around $1.2\sim 1.4$ GeV, the $a_2(1320)$ production provides the dominant contribution to the cross section. In our description, we do not include the contribution of the $\pi(1300)$ resonance~\cite{ARGUS:1996ith,L3:1997mpi}, because the parameters of such large width $\pi(1300)$ state  come with large uncertainties. Its inclusion, however, might account for some deviations in the total cross section observed at both sides of the $a_2(1320)$ resonance peak. In the energy region beyond $1.4$ GeV, the $f_2(1270)\pi^0$ production mechanism in the $t$- and $u$-channel starts contributing and becomes important beyond $1.6$ GeV. Its inclusion yields a rather good description of the total cross section in that region. Our results are more consistent with ARGUS data rather than L3 data. In our analysis, no indication is found for a significant contribution from the $\pi_2(1670)$ and $a_2(1700)$ resonances.

Furthermore, in Fig.~\ref{Fig:TotCross_lowE} we present our result of total cross section in the low energy region: $3m_\pi<W<0.9$ GeV. The results of the LO and NLO ChPT~\cite{Talavera:1995fx} are also shown for comparison. One can see that the $\gamma\gamma\to \pi^+\pi^-\pi^0$ cross section obtained at one loop in ChPT is significantly larger than the LO predictions. The rather small LO result is due to the drastic destructive interference between the $\pi^0$-pole and contact diagrams in the LO amplitude, as stated in Ref.~\cite{Talavera:1995fx}. Our prediction is also larger than the NLO cross section of this energy region. Our amplitude contains the physics of the $\gamma\gamma\to\sigma\pi$ channel, i.e. the physical $\sigma/f_0(500)$ contribution by taking into account the rescattering effect of the final $\pi^+\pi^-$ states through the Omn\`{e}s function. This leads to a significant enhancement as compared to the chiral calculations. Going to the energy region where the experimental data of L3 and ARGUS are available, our result is consistent with the L3 data by including the contributions of the $\sigma\pi^0$ and the $\rho\pi$ channels. Upon naive extrapolation, the chiral calculation seems to fall significantly below the data points around $0.8$ GeV. This likely indicates that such extrapolation is unreliable and that the range of validity of the chiral amplitude is much more limited for this process.

\begin{figure*}[htbp]
    \centering
    \includegraphics[width=0.9\textwidth]{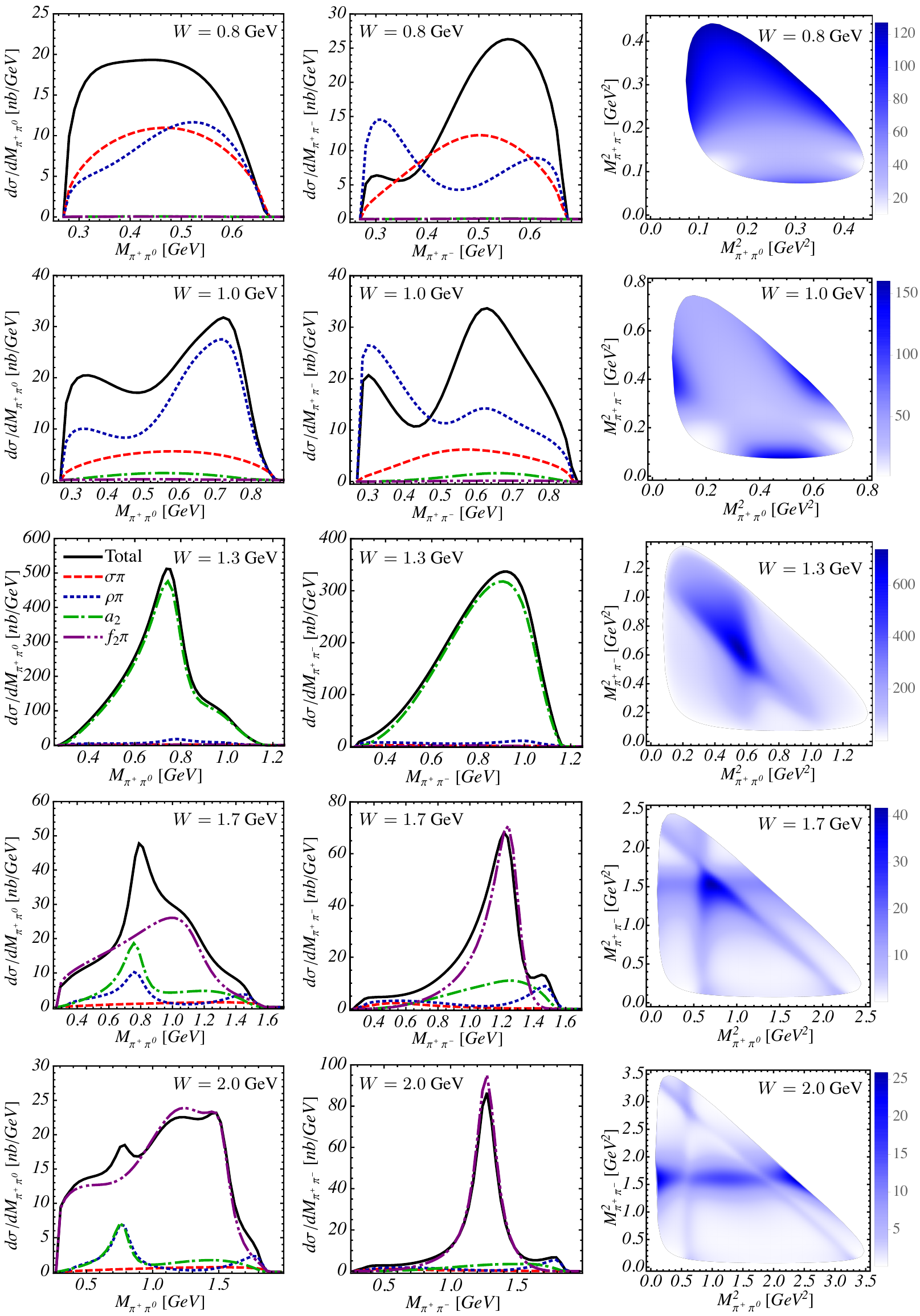}
    \caption{The predicted $M_{\pi^+\pi^0}$ and $M_{\pi^+\pi^-}$ mass distributions of $\gamma\gamma\to\pi^+\pi^-\pi^0$ and the Dalitz plot ($M_{\pi^+\pi^0}^2$ versus $M_{\pi^+\pi^-}^2$) for $W=0.8$, $1.0$, $1.3$, $1.7$, and $2.0$ GeV, respectively.  The black curves denote the total results of our model. The individual contributions from the $\sigma/f_0(500)\pi^0$, $\rho^\pm\pi^\mp$, $a_2(1320)$, and $f_2(1270)\pi^0$ channels are indicated by the dashed (red), dotted (blue), dash-dotted (green), and dash-double-dotted (purple) curves.}
    \label{Fig:DiffCross_prediction}
\end{figure*}

Next, we show the prediction of our model for the invariant mass distributions $d\sigma/dM_{\pi^+\pi^0}$ and $d\sigma/dM_{\pi^+\pi^-}$ and the Dalitz plot ($M_{\pi^+\pi^0}^2$ versus $M_{\pi^+\pi^-}^2$) for different $\gamma\gamma$ c.m. energies. In Fig.~\ref{Fig:DiffCross_prediction},  several values of the total energy ($W=0.8$, $1.0$, $1.3$, $1.7$ and $2.0$ GeV) are presented to cover the current and forthcoming experimental energy range.   
\begin{itemize}
\item The mass distributions at $W=0.8$ GeV are saturated by the $\sigma/f_0(500) \pi^0$ and $\rho\pi$ contributions. The behavior of the $\sigma/f_0(500)$ production channel resembles a phase space distribution. A similar phenomenon is observed in the $M_{\pi^+\pi^0}$ spectrum of the $\rho\pi$ channel, since the $\rho$ resonance cannot be produced on-shell at this energy. In the $M_{\pi^+\pi^-}$ distribution, the $\rho\pi$ contribution presents the typical behavior of the kinematic reflection of the $\rho\pi$ channel as shown in the Dalitz plot, which will be explained in detail in the following.  
\item Going to $W=1.0$ GeV, both distributions are dominated by the $\rho\pi$ channel. Although the contribution from $\sigma/f_0(500)\pi^0$ is relatively small, its interference with the $\rho\pi$ channel cannot be neglected. For the $\rho\pi$ distributions, besides a $\rho^+$ resonance peak clearly showing up in the $d\sigma/dM_{\pi^+\pi^0}$ distribution, an interesting observation is the increasing bump structures of both distributions at low invariant masses. This is due to the kinematic reflection of the production of $\rho^{+}$ and $\rho^-$ resonances. Such mechanism is clearly presented in the Dalitz plot: the spin-1 $\rho^+$ resonance gives two enhancements at the edges of the Daltiz plot with $M_{\pi^+\pi^0}^2\sim (m_\rho\pm \Gamma_\rho/2) ^2$, and the $\rho^-$ resonance produces the off-diagonal distribution at the edges of $M_{\pi^+\pi^0}^2\sim (m_\rho\pm \Gamma_\rho/2) ^2$ and $M_{\pi^+\pi^-}^2\sim (m_\rho\pm \Gamma_\rho/2) ^2$. Furthermore, the broad peak in $d\sigma/dM_{\pi^+\pi^-}$ is mainly due to the constructive interference of $\rho\pi$ and $\sigma/f_0(500)$ channels. 
\item At $W=1.3$ GeV, the $s$-channel of $a_2(1320)$ production plays the leading role for both mass distributions. The $\rho\pi$ channel has a relatively small contribution. The $\rho^+$ resonance peak is clearly seen in the $M_{\pi^+\pi^0}$ distribution because of the intermediate decay mode of $a_2(1320)\to\rho^\pm\pi^\mp$. The Dalitz plot also shows the bands due to the $\rho^\pm$ states. Then, the integration over $M_{\pi^+\pi^0}$ leads to a broad peak at $d\sigma/dM_{\pi^+\pi^-}$ around $M_{\pi^+\pi^-}=0.9$ GeV. 

\item For total energy of $W=1.7$ GeV, the contribution of the $f_2(1270) \pi^0$ channel is dominant. Through the interplay between the $\rho^\pm \pi^\mp$, $a_2(1320)$, and $f_2(1270)\pi^0$ channels, the mass distributions present several characteristic features: the asymmetric shapes with enhancements at both end points of $M_{\pi^+\pi^0}$ and $M_{\pi^+\pi^-}$. For the $M_{\pi^+\pi^0}$ spectrum, the $\rho$ resonance peak is enhanced by the constructive interference between $\rho\pi$ and $a_2(1320)$ [$f_2(1270)\pi^0$] channels in combination with the destructive interference between $a_2(1320)$ and $f_2(1270)\pi^0$ channels. A notable shoulder beyond the $\rho$ peak is mainly due to the $f_2(1270)\pi^0$ contribution. At small $M_{\pi^+\pi^0}$ value, the shape is determined by $f_2(1270)\pi^0$ channel, while at large $M_{\pi^+\pi^0}$ value, the spectrum is enhanced by the $\rho\pi$ channel, which originates from the kinematic reflection of the $\rho$ resonance production. For the $M_{\pi^+\pi^-}$ distribution, the $f_2(1270)$ resonance peak is prominent, which is not 
affected by the destructive interference with the $a_2(1320)$ channel. The enhancement at both edges of the $M_{\pi^+\pi^-}$ distribution is mainly from the $\rho\pi$ channel, which has the ``two-peak'' structure due to the kinematic reflection. A similar observation of the $\rho\pi$ channel was recently found by Belle II collaboration in the study of the $B^+\to \pi^+\pi^0\pi^0$ process~\cite{Belle:2022dgi}.
\item At $W=2$ GeV, the role of the $f_2(1270)$ channel is the main feature in both distributions, which is also represented in the Dalitz plot. In addition, the small $\rho$ peak is shown in the $M_{\pi^+\pi^0}$ spectrum. For the $M_{\pi^+\pi^-}$ distribution, besides the pronounced $f_2(1270)$ peak, there is also a visible enhancement at the largest $M_{\pi^+\pi^-}$ due to the edge bump of the $\rho\pi$ channel. 
\end{itemize}
In summary, the evolution with energy of the Dalitz plots in Fig.~\ref{Fig:DiffCross_prediction} demonstrates the different underlying physical mechanisms in the energy range from 0.8 GeV to 2.0 GeV. It can be probed in more detail by the forthcoming experiment data from BESIII.

\begin{figure*}[htbp]
    \centering
    \begin{tabular}{c c}
        \textbf{ARGUS} & \textbf{L3} \\[0.5em]
 \includegraphics[width=0.45\textwidth]{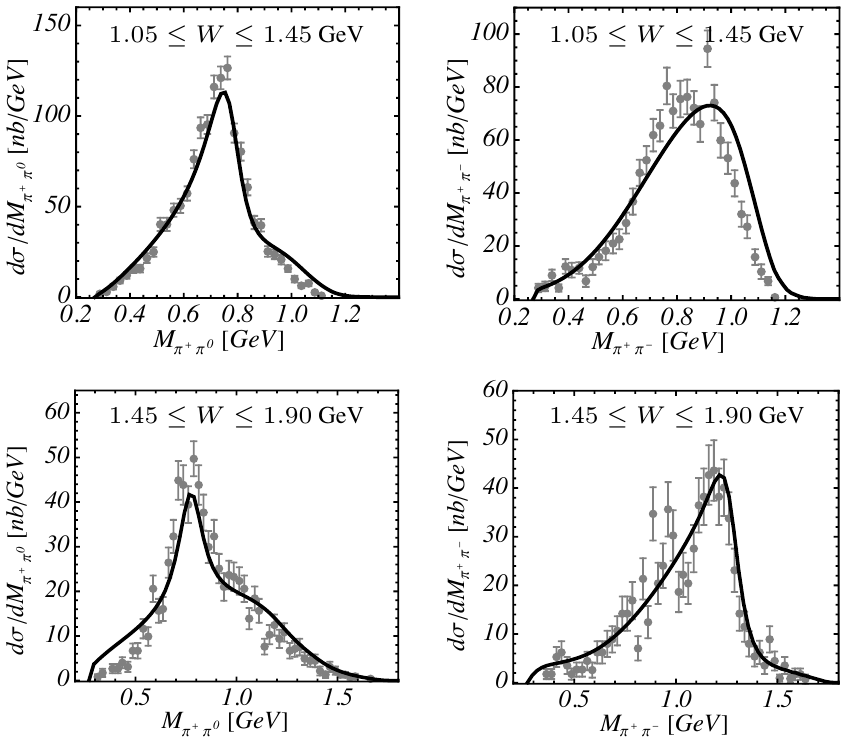}~~~ & ~~~
    \includegraphics[width=0.45\textwidth]{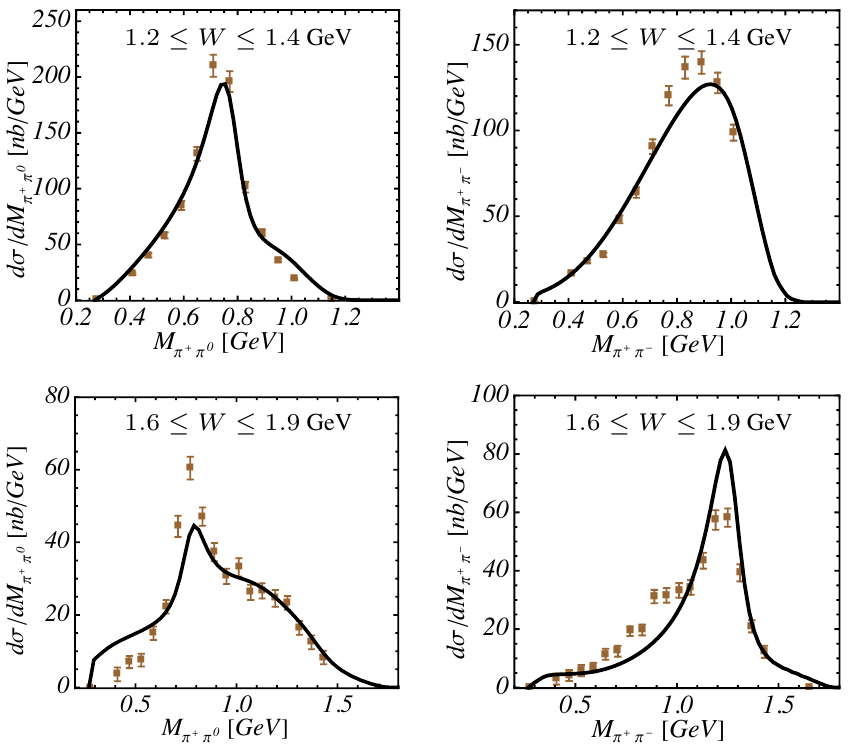}
    \end{tabular}
    \caption{The comparison results of $M_{\pi^+\pi^0}$ and $M_{\pi^+\pi^-}$ mass distributions at different energy intervals. The black curves represent the weighted average results of our model. 
    The normalized data points of ARGUS~\cite{ARGUS:1996ith}  and L3~\cite{Shchegelsky:2006es} collaborations are denoted by the gray dots (brown squares).}
    \label{Fig:DiffCross_data}
\end{figure*}

Finally, we present the description of the invariant mass distributions $d\sigma/dM_{\pi^+\pi^0}$ and $d\sigma/dM_{\pi^+\pi^-}$ as compared to the ARGUS and L3 data. The existing experimental data have unfortunately low statistics, particularly for the L3 results~\cite{Shchegelsky:2006es}. As a result, both measurements do not report the $M_{\pi^+\pi^0}$ and $M_{\pi^+\pi^-}$ distributions at a single total energy but within some energy intervals: ARGUS gives the mass distributions in Fig.~6 of Ref.~\cite{ARGUS:1996ith} within two intervals, $1.05\leq W \leq 1.45$ GeV and $1.45 \leq W\leq 1.90$ GeV; while L3 reports the data within the energy intervals, $1.2\leq W \leq 1.4$ GeV and $1.6 \leq W\leq 1.9$ GeV, as shown in Fig.~6 and 7 of Ref.~\cite{Shchegelsky:2006es}. Furthermore, the angular distribution and efficiency of their detectors are unknown, which causes difficulty to perform an exact comparison. 

Therefore, in order to carry out a meaningful comparison, we first perform the weighted average of our theoretical prediction of the invariant mass distributions via the following formula: 
\begin{equation}
    \left(\frac{d\sigma}{d M_{\pi^+\pi^{0,-}}}\right)_\mathrm{avg} = \frac{1}{n}\sum\limits_{i=1}^{n} \frac{\sigma_i}{\sigma_\mathrm{max}}\, \frac{d\sigma_i}{d M_{\pi^+\pi^{0,-}}},
\end{equation}
in each energy interval $W_\mathrm{min}\leq W\leq W_\mathrm{max}$. The number of the selected points $n$ is $n=1+(W_\mathrm{max}-W_\mathrm{min})/\Delta W$, where the energy step is taken as $\Delta W=0.5$ GeV according to the bins of the total cross section of ARGUS and L3. The weighting factor of the invariant mass distribution, $d\sigma_i/dM_{\pi^+\pi^{0,-}}$, at $W=W_i$ is chosen $\sigma_i/\sigma_\mathrm{max}$, where $\sigma_i\equiv \sigma(W_i=W_\mathrm{min}+(i-1)\Delta W),~i=1,...,n$ is the total cross section, and $\sigma_\mathrm{max}$ is the maximum value of total cross section among the selected points over the corresponding energy interval. Next, for the experimental distributions, we normalize the ARGUS and L3 data to generate the same area as our averaged mass distributions in the corresponding energy intervals. 

The comparison of the theoretical invariant mass distributions $d\sigma/dM_{\pi^+\pi^0}$ and $d\sigma/dM_{\pi^+\pi^-}$ with the experimental ones is shown in Fig.~\ref{Fig:DiffCross_data}.
We notice that the shapes of our prediction are globally very consistent with the ARGUS data in both energy intervals and with the L3 data in the low-energy interval. Note that there are no reported data points of L3 $d\sigma/d M_{\pi^+\pi^-}$ distribution of $M_{\pi^+\pi^-}$ above $1$ GeV with $1.2\leq W \leq 1.4$ GeV. A relatively large deviation from our predicted distributions is observed at the high-energy interval of L3 data. This is mainly due to the difference which is observed between both data sets in the total cross section from 1.6 to 1.8 GeV (Fig.~\ref{Fig:TotCross}). Besides, there are some differences between our results and ARGUS data, such as the $d\sigma/d M_{\pi^+\pi^-}$ distribution at the low-energy interval, the $d\sigma/d M_{\pi^+\pi^0}$ distribution at very small $M_{\pi^+\pi^-}$ with  $1.45\leq W \leq 1.90$ GeV. 
Although, the data comparison shows that our model captures the qualitative features of the data, it also clearly calls for high statistics data  to refine the theoretical analysis.

\section{Conclusion and perspectives}\label{SecIV}
We constructed a theoretical model for the $\gamma\gamma\to\pi^+\pi^-\pi^0$ process by including the contributions of the $a_2(1320)$ resonance excitation, as well as the contributions of the $\sigma/f_0(500)\pi^0$, $f_2(1270)\pi^0$, and $\rho^\pm(770)\pi^\mp$ channels. To cover the interested energy range from $0.8$ GeV to $2$ GeV, the $\rho$ and $\pi$ exchange mechanisms were Reggezied to achieve a good high-energy behavior. As a proof of the applicability, we analyzed the current experimental data of total cross section and invariant mass distributions from ARGUS and L3 Collaborations and found a rather good description. In particular, our model favors the smaller total cross section values of the L3 data at low energies. Furthermore, we also present the theoretical predictions for the total energies $W=0.8$, $1.0$, $1.3$, $1.7$, and $2.0$ GeV, which will be investigated by the forthcoming BESIII measurements of the $\gamma\gamma \to\pi^+\pi^-\pi^0$ reaction. 
Such renewed experimental effort is needed to clarify the existing data situation on the $\gamma\gamma \to \pi^+\pi^-\pi^0$ reaction, as well as its extensions to single- and double-virtual photon fusion processes. On the theoretical side, one needs to improve the current model by corroborating the dispersion theory for this $2\to3$ process, extending the success of the dispersive approach in the $\gamma^{(*)}\gamma^{(*)} \to \pi\pi$ reaction \cite{GarciaMartin:2010cw,Hoferichter:2011wk,Moussallam:2013una,Danilkin:2018qfn,Hoferichter:2019nlq,Danilkin:2019opj}. Our work may serve as first step towards a data-driven approach for the $\gamma^{(*)}\gamma^{(*)}\to\pi^+\pi^-\pi^0$ reaction, which is necessary to achieve a controllable estimate of the hadronic light-by-light contribution to $(g-2)_\mu$ with the three-pion intermediate state. 

\begin{acknowledgments} 
This work was supported by the Deutsche Forschungsgemeinschaft (DFG, German Research Foundation), in part through the Research Unit [Photon-photon interactions in the Standard Model and beyond, Projektnummer 458854507 - FOR 5327], and in part through the Cluster of Excellence [Precision Physics, Fundamental Interactions, and Structure of Matter] (PRISMA$^+$ EXC 2118/1) within the German Excellence Strategy (Project ID 39083149).
\end{acknowledgments}

\begin{widetext}
\appendix*
\section{Lorentz decomposition of the $\gamma\gamma\to V P$ reaction}\label{App}
We present below the general Lorentz decomposition for the scattering amplitude of two photons fusion to a pseudo-scalar and a vector meson. Taking into account the crossing symmetry, the on-shell condition of final vector meson and the Schouten identity, we found 6 independent tensor structures for the $\gamma(k_1)\gamma(k_2)\to V(p_1) P(p_2)$ reaction, 
\begin{equation}\label{Eq:tensorbasis}
\begin{aligned}
T^1_{\mu\nu,\alpha}(k_1,k_2,\Delta) &= (k_1-k_2)_\alpha \,\epsilon_{\mu\nu\gamma\beta} {k_1}^\gamma {k_2}^\beta, \\
T^2_{\mu\nu,\alpha}(k_1,k_2,\Delta)  &= (k_1+k_2)_\alpha \,\epsilon_{\mu\nu\gamma\beta} {k_1}^\gamma {k_2}^\beta, \\
T^{3}_{\mu\nu,\alpha}(k_1,k_2,\Delta)  &= \left[ g_{\mu\nu} (k_1\cdot k_2) - {k_1}_\mu {k_2}_\nu - {k_1}_\nu {k_2}_\mu \right] \epsilon_{\alpha \sigma\gamma\beta} {k_1}^\sigma {k_2}^\gamma \Delta^\beta,\\
 T^{4}_{\mu\nu,\alpha}(k_1,k_2,\Delta)  &= -\left[  (k_2\cdot\Delta){k_1}_\mu +(k_1\cdot\Delta) {k_2}_\mu -(k_1\cdot k_2)\Delta_\mu \right] 
\epsilon_{\alpha\nu\gamma\beta}{k}^\gamma{k_2}^\beta \\
&\quad - \left[(k_2\cdot\Delta){k_1}_\nu  +(k_1\cdot\Delta) {k_2}_\nu -(k_1\cdot k_2)\Delta_\nu \right] \epsilon_{\alpha\mu\gamma\beta}{k}^\gamma{k_2}^\beta, \\
 T^{5}_{\mu\nu,\alpha}(k_1,k_2,\Delta)  &= 
  \left[  (k_2\cdot\Delta)  {k_1}_\mu + (k_1\cdot\Delta) {k_2}_\mu - (k_1\cdot k_2)  \Delta_\mu\right]\, {k_1}_\alpha\,  \epsilon_{\nu\sigma\gamma\beta} {k_1}^\sigma {k_2}^\gamma \Delta^\beta \\
  &\quad - \left[ (k_2\cdot\Delta) {k_1}_\nu + (k_1\cdot\Delta)  {k_2}_\nu - (k_1\cdot k_2) \Delta_\nu\right]\,  {k_2}_\alpha\,  \epsilon_{\mu\sigma\gamma\beta}{k_1}^\sigma {k_2}^\gamma \Delta^\beta, \\
T^{6}_{\mu\nu,\alpha}(k_1,k_2,\Delta)  &= -
  \left[  (k_2\cdot\Delta)  {k_1}_\mu + (k_1\cdot\Delta)  {k_2}_\mu - (k_1\cdot k_2)  \Delta_\mu\right] \, {k_1}_\alpha\, \epsilon_{\nu\sigma\gamma\beta} {k_1}^\sigma {k_2}^\gamma \Delta^\beta \\
  &\quad - \left[ (k_2\cdot\Delta)  {k_1}_\nu + (k_1\cdot\Delta)  {k_2}_\nu - (k_1\cdot k_2)  \Delta_\nu\right] \, {k_2}_\alpha \, \epsilon_{\mu\sigma\gamma\beta}{k_1}^\sigma {k_2}^\gamma \Delta^\beta,
\end{aligned}
\end{equation}
with the momentum difference $\Delta=p_1-p_2$. 

We applied these tensor structures to decompose the $\gamma\gamma\to\rho^\pm\pi^\mp$ amplitude as shown in Fig.~\ref{Fig:FeynALL}(d-f), and found the corresponding scalar functions:  
\begin{align}
   F_1^{\gamma\gamma \to \rho^+\pi^-}(s,t,u) &=   -F_1^{\gamma\gamma \to \rho^-\pi^+}(s,t,u) \nonumber\\
   &=\frac{e^2  g_{\rho\pi\gamma}}{4\, m_\pi\, s^2} \Biggl\{\left(  2 \left(m_\pi^2-m_\rho^2\right)^2 -s\left(m_\pi^2+7 m_\rho^2\right) +  s^2 \right)
   \left[ \frac{1}{t-m_\rho^2} - \frac{1}{u-m_\rho^2} \right] \nonumber \\
   &\quad - 2\left(\left(m_\pi^2-m_\rho^2\right)^2-s\left(3 m_\pi^2+ m_\rho^2\right)\right) \left[ \frac{1}{t-m_\pi^2} -\frac{1}{u-m_\pi^2}\right]\Biggr\}, \label{Eq:scalarfuns2} \\
  F_2^{\gamma\gamma \to \rho^+\pi^-}(s,t,u) &=- F_2^{\gamma\gamma \to \rho^-\pi^+}(s,t,u)  \nonumber \\
  &=\frac{e^2  g_{\rho \pi \gamma } }{4\, m_\pi\, s^2}  \Biggl\{ \left(2\left(m_{\pi }^2-m_{\rho }^2\right)^2-s
   \left(m_{\pi }^2-9 m_{\rho }^2\right)+s^2\right)\left[ \frac{1}{t-m_{\rho}^2} + \frac{1}{u-m_{\rho}^2} \right]  \nonumber \\
   & \quad + 2\left(\left(m_{\pi}^2-m_{\rho }^2\right)^2+s \left(3 m_{\pi }^2+m_{\rho}^2\right)\right) \left[\frac{1}{t-m_{\pi}^2} +\frac{1}{u-m_{\pi}^2}  \right] +4s \Biggr\},\\
    F_3^{\gamma\gamma \to \rho^+\pi^-}(s,t,u) &=-F_3^{\gamma\gamma \to \rho^-\pi^+}(s,t,u) \nonumber \\
    &= \frac{e^2  g_{\rho\pi\gamma}}{s \, m_\pi} 
    \left[\frac{1}{t-m_\rho^2} - \frac{1}{u-m_\rho^2}\right],\\
  F_4^{\gamma\gamma \to \rho^+\pi^-}(s,t,u) &= -F_4^{\gamma\gamma \to \rho^-\pi^+}(s,t,u) \nonumber \\
  &=- \frac{e^2 g_{\rho \pi \gamma }}{2 m_{\pi } s^2} \Biggl\{ \left(s+2m_{\rho}^2-2 m_{\pi}^2\right)\left[\frac{1}{t-m_{\rho}^2}-\frac{1}{u-m_{\rho}^2}\right] \nonumber \\
   &\quad + 2(m_{\pi}^2 - m_{\rho}^2)\left[\frac{1}{t-m_{\pi}^2}-\frac{1}{u-m_{\pi}^2}\right] \Biggr\},\\
 F_5^{\gamma\gamma \to \rho^+\pi^-}(s,t,u)  &= -F_5^{\gamma\gamma \to \rho^-\pi^+}(s,t,u) \nonumber \\
 &=-\frac{e^2  g_{\rho \pi \gamma }}{m_{\pi } s^2} \left[\frac{1}{t-m_{\rho }^2} +\frac{1}{u-m_{\rho }^2} + \frac{1}{t-m_{\pi }^2}
 +\frac{1}{u-m_{\pi }^2}\right], \\
 F_6^{\gamma\gamma \to \rho^+\pi^-}(s,t,u) &= -F_6^{\gamma\gamma \to \rho^-\pi^+}(s,t,u) \nonumber \\
  &=- \frac{e^2 g_{\rho \pi \gamma }
   }{m_{\pi } s^2} \left[\frac{1}{t-m_{\rho}^2}-\frac{1}{u-m_{\rho }^2} 
   -\frac{1}{t-m_{\pi}^2}+\frac{1}{u-m_{\pi}^2}\right] \label{Eq:scalarfuns7} .
\end{align}

 \end{widetext}
 
\bibliographystyle{apsrev4-1}
\bibliography{ggTo3pions}

\end{document}